\documentclass[fleqn,twoside]{article}
\usepackage[headings]{espcrc2}

\readRCS
$Id: espcrc2.tex,v 1.2 2004/02/24 11:22:11 spepping Exp $
\usepackage{epsfig}

\def\d{{\rm d}}

\def\e{\epsilon}

\title{
Antenna subtraction method for jet calculations at NNLO}

\author{A.\ Gehrmann-De Ridder,
\address{Institute for Theoretical Physics, ETH, CH-8093 Z\"urich,
Switzerland}
T.\ Gehrmann,
\address{Institut f\"ur Theoretische Physik,
Universit\"at Z\"urich, CH-8057 Z\"urich, Switzerland}
E.W.N.\ Glover
\address{Institute of Particle Physics Phenomenology,
        University of Durham, Durham, DH1 3LE, UK}}

\begin{document}
\unitlength 1cm

\begin{abstract}
We describe the antenna subtraction method for treating real emission 
singularities in the calculation of jet observables at NNLO accuracy, 
in particular in view of the computation $e^+e^- \to 3$~jets at NNLO.
\end{abstract}
\maketitle

\thispagestyle{myheadings}
\markright{ZU-TH 04/06, IPPP/06/04}

\section{Introduction} 
Using experimental data on jet production observables to extract 
QCD parameters, especially the QCD coupling constant $\alpha_s$, one 
finds that the dominant source of error on these parameters is 
very often the 
uncertainty inherent to the QCD calculations of the obervables. 
These calculations are at present available to next-to-leading
(NLO) accuracy, their precision can be improved only by calculating
corrections to the next perturbative order: next-to-next-to-leading order
(NNLO).

In the recent past, many ingredients to NNLO calculations of collider
 observables have been derived (see~\cite{ourant} and references therein): 
the massless
two-loop $2\to 2$ and $1 \to 3$ matrix elements relevant to 
NNLO jet production 
have been computed
and are now available for many processes of phenomenological relevance.
The one-loop corrections to $2\to 3$ and $1\to 4$ matrix elements 
have been known for longer and form part 
of NLO calculations of 
the respective multi-jet observables. 
These NLO matrix elements naturally
contribute to NNLO jet observables of lower multiplicity if one of the 
partons involved becomes unresolved (soft or collinear). 
In these cases, the infrared 
singular parts of the matrix elements need to be extracted and integrated 
over the phase space appropriate to the unresolved configuration 
to make the infrared pole structure explicit. 
As a final ingredient, the tree level $2\to 4$ and $1\to 5$ processes also
contribute to ($2\to 2$)- and ($1 \to 3$)-type jet observables at NNLO. 
These contain double real radiation singularities corresponding to two 
partons becoming simultaneously soft and/or 
collinear. To compute the contributions from single unresolved radiation 
at one-loop and double real radiation at tree level,  one has to 
find subtraction terms which coincide with the full matrix elements 
in the unresolved limits 
and are still sufficiently simple to be integrated analytically in order 
to cancel  their  infrared pole structure with the two-loop virtual 
 contributions. It is the aim of this talk to present a systematic 
method, named antenna subtraction, to construct NNLO subtraction
terms and to illustrate its applications.

\section{Antenna subtraction}
\begin{figure}[t] 
\epsfig{file=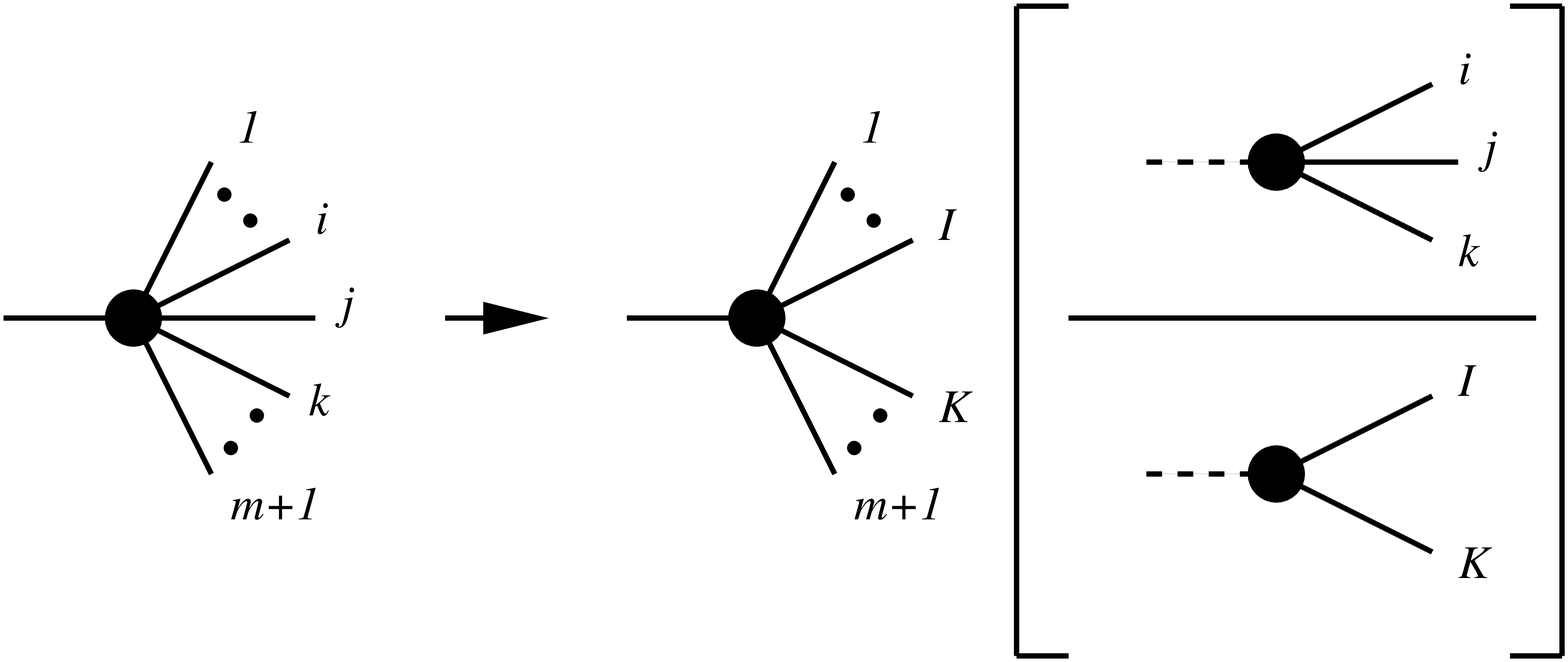,height=2.9cm}
\vspace{-5mm}
\caption{Illustration of NLO antenna factorisation representing the
factorisation of both the squared matrix elements and the 
$(m+1)$-particle phase
space. The term in square brackets
represents both the antenna function and the antenna phase space.
\label{fig:nloant}
}
\vspace{2mm}
\epsfig{file=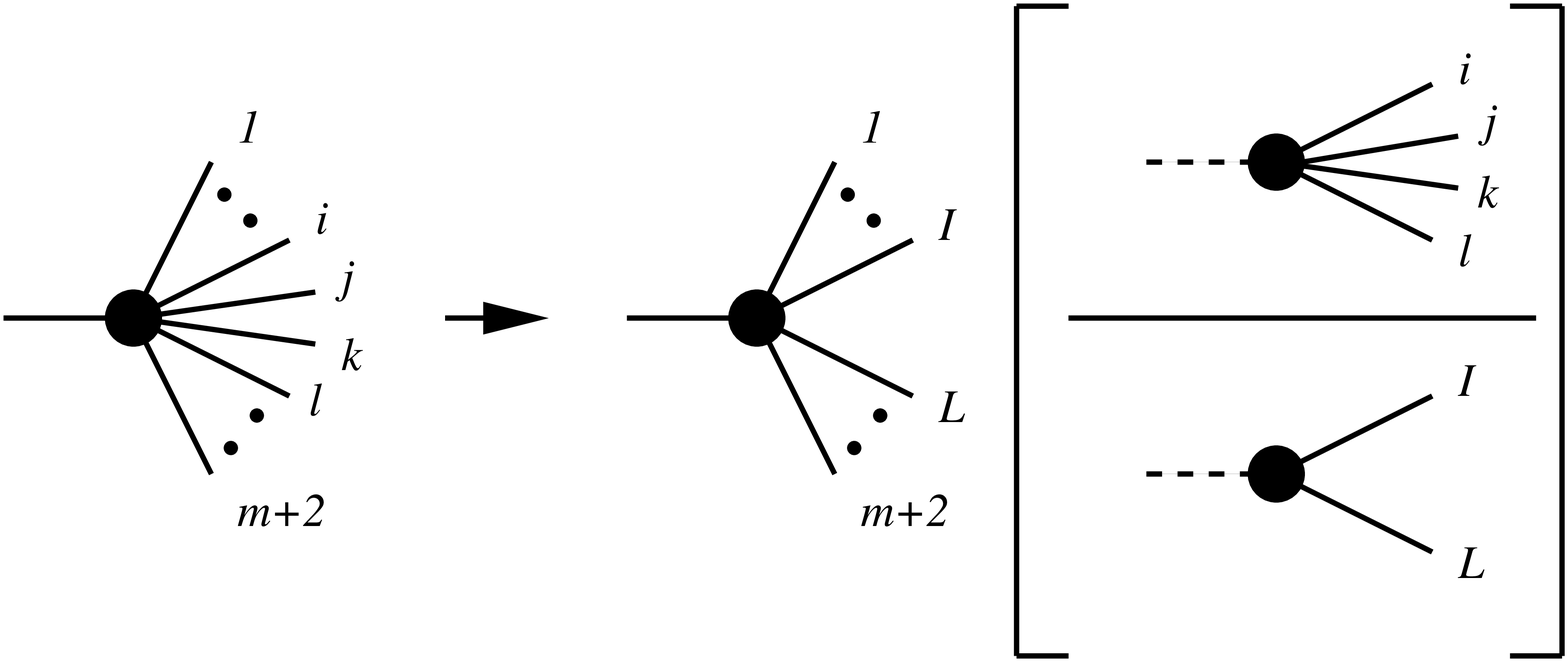,,height=2.9cm}
\vspace{-5mm}
\caption{\label{fig:sub2a} Illustration 
of NNLO antenna factorisation representing the
factorisation of both the squared matrix elements and the $(m+2)$-particle 
phase
space when the unresolved particles are colour connected. 
The term in square brackets
represents antenna function and phase space.}
\end{figure}
An $m$-jet cross section at NLO is obtained by summing contributions from 
$(m+1)$-parton tree level and $m$-parton one-loop processes:
\begin{eqnarray*}
\lefteqn{{\rm d}\sigma_{NLO}=\int_{{\rm d}\Phi_{m+1}}\left({\rm d}\sigma^{R}_{NLO}
-{\rm d}\sigma^{S}_{NLO}\right)} \nonumber \\
&&
+\left [\int_{{\rm d}\Phi_{m+1}}
{\rm d}\sigma^{S}_{NLO}+\int_{{\rm d}\Phi_{m}}{\rm d}\sigma^{V}_{NLO}\right].
\end{eqnarray*}
The cross section ${\rm d}\sigma^{R}_{NLO}$ is the $(m+1)$-parton tree-level
cross section, 
while 
${\rm d}\sigma^{V}_{NLO}$ is the one-loop virtual correction to the 
$m$-parton Born cross section ${\rm d}\sigma^{B}$. Both contain infrared 
singularities, which are explicit poles in $1/\e$ in ${\rm d}\sigma^{V}_{NLO}$,
while becoming explicit in ${\rm d}\sigma^{R}_{NLO}$ only after integration 
over the phase space.  In general, this integration  involves the
(often iterative) definition of the jet observable, such that  an analytic
integration is not feasible (and also not appropriate). Instead,   one would
like to have a flexible method that can be easily adapted to  different jet
observables or jet definitions. Therefore, the infrared singularities  
of the real radiation
contributions should be extracted using  infrared subtraction  terms.
One introduces ${\rm d}\sigma^{S}_{NLO}$, which is a counter-term for  
 ${\rm d}\sigma^{R}_{NLO}$, having the same unintegrated
singular behaviour as ${\rm d}\sigma^{R}_{NLO}$ in all appropriate limits.
Their difference is free of divergences 
and can be integrated over the $(m+1)$-parton phase space numerically.
The subtraction term  ${\rm d}\sigma^{S}_{NLO}$ has 
to be integrated analytically over all singular regions of the 
$(m+1)$-parton phase space. 
The resulting cross section added to the virtual contribution 
yields an infrared finite result. 
Several methods for  constructing
 NLO subtraction terms systematically  were proposed in the 
literature~\cite{cullen,ant,cs,singleun}. For some of these methods, 
extension to NNLO was discussed~\cite{nnlosub} and partly worked out. 
We focus on the  antenna subtraction method~\cite{cullen,ant}, 
which we extend to NNLO. 

The basic idea of the antenna subtraction approach at NLO is to construct 
the subtraction term  
${\rm d}\sigma^{S}_{NLO}$
from antenna functions. Each antenna function encapsulates 
all singular limits due to the 
 emission of one unresolved parton between two colour-connected hard
partons (tree-level three-parton antenna function).
This construction exploits the universal factorisation of 
phase space and squared matrix elements in all unresolved limits,
depicted in Figure~\ref{fig:nloant}.
The individual antenna functions are obtained by normalising 
three-parton tree-level matrix elements to the corresponding two-parton 
tree-level matrix elements. 

At NNLO, the $m$-jet production is induced by final states containing up to
$(m+2)$ partons, including the one-loop virtual corrections to $(m+1)$-parton final 
states. As at NLO, one has to introduce subtraction terms for the 
$(m+1)$- and $(m+2)$-parton contributions. 
Schematically the NNLO $m$-jet cross section reads,
\begin{eqnarray*}
\lefteqn{{\rm d}\sigma_{NNLO}=\int_{{\rm d}\Phi_{m+2}}\left({\rm d}\sigma^{R}_{NNLO}
-{\rm d}\sigma^{S}_{NNLO}\right) }\nonumber \\ 
&&+\int_{{\rm d}\Phi_{m+1}}\left({\rm d}\sigma^{V,1}_{NNLO}
-{\rm d}\sigma^{VS,1}_{NNLO}\right)\nonumber \\&&
+ \int_{{\rm d}\Phi_{m+2}}
{\rm d}\sigma^{S}_{NNLO}
+\int_{{\rm d}\Phi_{m+1}}{\rm d}\sigma^{VS,1}_{NNLO}  \nonumber \\ 
&&
+ \int_{{\rm d}\Phi_{m}}{\rm d}\sigma^{V,2}_{NNLO}\;,
\end{eqnarray*}
where $\d \sigma^{S}_{NNLO}$ denotes the real radiation subtraction term 
coinciding with the $(m+2)$-parton tree level cross section 
 $\d \sigma^{R}_{NNLO}$ in all singular limits~\cite{doubleun}. 
Likewise, $\d \sigma^{VS,1}_{NNLO}$
is the one-loop virtual subtraction term 
coinciding with the one-loop $(m+1)$-parton cross section 
 $\d \sigma^{V,1}_{NNLO}$ in all singular limits~\cite{onelstr}. 
Finally, the two-loop correction 
to the $m$-parton cross section is denoted by ${\rm d}\sigma^{V,2}_{NNLO}$.
\begin{figure*}[t]
\begin{center}
\epsfig{file=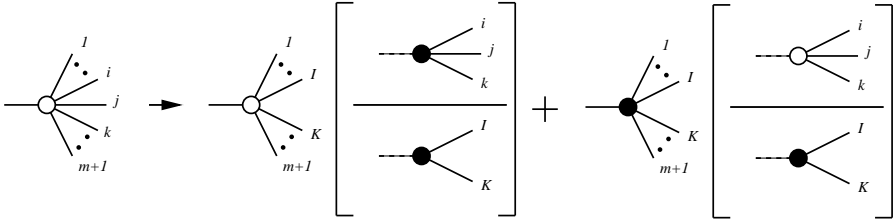,height=2.9cm}
\vspace{-9mm}
\end{center}
\caption{Illustration of NNLO antenna factorisation representing the
factorisation of both the one-loop
``squared" matrix elements (represented by the white blob)
and the $(m+1)$-particle phase
space when the unresolved particles are colour connected. 
\label{fig:subv}}
\end{figure*}

Both types of NNLO subtraction terms can be constructed from antenna 
functions. In ${\rm d}\sigma^{S}_{NNLO}$, we have to distinguish four
different types of unresolved configurations:
(a) One unresolved parton but the experimental observable selects only
$m$ jets;
(b) Two colour-connected unresolved partons (colour-connected);
(c) Two unresolved partons that are not colour connected but share a common
radiator (almost colour-unconnected);
(d) Two unresolved partons that are well separated from each other 
in the colour 
chain (colour-unconnected). Among those, configuration (a) is properly 
accounted for by a single tree-level three-parton antenna function 
like used already at NLO. Configuration (b) requires a 
tree-level four-parton antenna function (two unresolved partons emitted 
between a pair of hard partons) 
as shown in Figure~\ref{fig:sub2a}, while (c) and (d) are accounted for by 
products of two tree-level three-parton antenna functions. 

In single unresolved limits, the one-loop cross section 
$\d \sigma^{V,1}_{NNLO}$ is described by the sum of two terms~\cite{onelstr}: 
a tree-level splitting function times a one-loop cross section 
and a one-loop splitting function times a tree-level cross section. 
Consequently, the 
one-loop single unresolved subtraction term $\d \sigma^{VS,1}_{NNLO}$
is constructed from tree-level and one-loop three-parton antenna functions,
as sketched in Figure~\ref{fig:subv}. Several other terms in  
 $\d \sigma^{VS,1}_{NNLO}$ cancel with the results
from the integration of terms in 
the double real radiation subtraction term  $\d \sigma^{S}_{NNLO}$
over the phase space appropriate to one of the unresolved partons, thus 
ensuring the cancellation of all explicit infrared poles in the difference 
$\d \sigma^{V,1}_{NNLO}-\d \sigma^{VS,1}_{NNLO}$.

Finally, all remaining terms in 
$\d \sigma^{S}_{NNLO}$ and $\d \sigma^{VS,1}_{NNLO}$ have to be integrated 
over the four-parton and three-parton antenna phase spaces. After   
integration, the infrared poles are rendered explicit and
cancel with the 
infrared pole terms in the two-loop squared matrix element 
$\d \sigma^{V,2}_{NNLO}$. 

\section{Derivation of antenna functions}
The  subtraction terms $\d \sigma^{S}_{NLO}$,
$\d \sigma^{S}_{NNLO}$ 
and $\d \sigma^{VS,1}_{NNLO}$ require three different types of 
antenna functions corresponding to the different pairs of hard partons 
forming the antenna: quark-antiquark, quark-gluon and gluon-gluon antenna 
functions. In the past~\cite{cullen,ant}, NLO antenna functions were 
constructed by imposing definite properties in 
all single unresolved limits (two collinear limits 
and one soft limit for each 
antenna). 
This procedure turns out to be impractical at NNLO, where each antenna 
function must have definite behaviours in a large number of single and 
double unresolved limits. Instead, we derive these antenna functions in 
a systematic manner from physical matrix elements known to possess the 
correct limits. The quark-antiquark antenna functions can be obtained 
directly from 
the $e^+e^- \to 2j$ real radiation corrections at NLO and NNLO~\cite{our2j}. 
For quark-gluon and gluon-gluon antenna functions, effective Lagrangians 
are used to obtain tree-level processes yielding a quark-gluon or 
gluon-gluon final state. The antenna functions are then obtained from 
the real radiation corrections to these processes. 
Quark-gluon antenna functions 
were derived~\cite{chi} from the purely QCD 
(i.e.\ non-supersymmetric) NLO and NNLO corrections to the decay of 
a heavy neutralino into a massless gluino plus partons~\cite{hw}, while 
gluon-gluon antenna functions~\cite{h} result from the QCD corrections 
to Higgs boson decay into partons~\cite{hgg}. 

All tree-level three-parton and four-parton antenna functions 
and  three-parton one-loop antenna functions are listed in~\cite{ourant}, 
where we also integrate them,  using the 
phase space integration techniques described in~\cite{ggh}.

\section{Application to $e^+e^- \to 3$~jets}
To illustrate the application of antenna subtraction  on a 
non-trivial example,  we derived
in~\cite{ourant,CFsquare} the $1/N^2$-contribution to 
the NNLO corrections to  $e^+e^- \to 3$~jets. This colour factor receives
contributions from 
$\gamma^*\to q\bar q ggg$ and $\gamma^*\to q\bar q q\bar qg$
at tree-level~\cite{tree5p},  
$\gamma^*\to q\bar q gg$ and $\gamma^*\to q\bar q q\bar q$  at
one-loop~\cite{onel4p} 
and $\gamma^*\to q\bar q g$ at two-loops~\cite{twol3p}. 
The four-parton and 
five-parton final states contain infrared singularities, which 
are extracted using the antenna subtraction formalism.

In this contribution, all gluons are effectively photon-like, and couple 
only to the quarks, but not to each other. Consequently, only quark-antiquark
antenna functions appear in the construction of the subtraction terms. 

Starting from the program {\tt EERAD2}~\cite{cullen}, which computes 
the four-jet
production at NLO, we implemented the NNLO antenna subtraction method 
for the $1/N^2$ colour factor contribution to $e^+e^-\to 3j$. {\tt EERAD2}
already  contains the five-parton and four-parton 
matrix elements relevant here, as well as the NLO-type subtraction terms. 

The implementation contains three channels, classified 
by their partonic multiplicity: 
(a) in the five-parton channel, we
integrate ${\rm d}\sigma_{NNLO}^{R} - {\rm d}\sigma_{NNLO}^{S}$;
(b) in the four-parton channel, we integrate
${\rm d}\sigma_{NNLO}^{V,1} - {\rm d}\sigma_{NNLO}^{VS,1}$;
(c) in the three-parton channel, we integrate
${\rm d}\sigma_{NNLO}^{V,2} +{\rm d}\sigma_{NNLO}^{S}
+ {\rm d}\sigma_{NNLO}^{VS,1}$.
The numerical integration over these channels is carried out by Monte Carlo 
methods. 

By construction, the integrands in the four-parton and 
three-parton channel are free of explicit infrared poles. In the 
five-parton and four-parton channel, we tested the proper implementation of 
the subtraction by generating trajectories of phase space points approaching 
a given single or double unresolved limit. 
Along these trajectories, we observe that the 
antenna subtraction terms converge locally towards the physical matrix 
elements, and that the cancellations among individual 
contributions to the subtraction terms take place as expected. 
Moreover, we checked the correctness of the 
subtraction by introducing a 
lower cut (slicing parameter) on the phase space variables, and observing 
that our results are independent of this cut (provided it is 
chosen small enough). This behaviour indicates that the 
subtraction terms ensure that the contribution of potentially singular 
regions of the final state phase space does not contribute to the numerical 
integrals, but is accounted for analytically. 

Finally, we noted in~\cite{ourant} that the 
infrared poles of the two-loop (including one-loop times one-loop) correction
to $\gamma^*\to q\bar qg$ are cancelled in all colour factors by a 
combination of integrated three-parton and four-parton
antenna functions.
This highly non-trivial cancellation 
clearly illustrates that the antenna functions derived 
here 
correctly approximate QCD matrix elements in all infrared singular limits at 
NNLO. They also outline the structure of infrared 
cancellations in $e^+e^-\to 3j$ at NNLO, and indicate the structure of the 
subtraction terms in all colour factors.

\section{Outlook} 

In this talk, 
we presented a new method for the subtraction of infrared 
singularities in the calculation of jet observables at NNLO. We introduced 
subtraction terms for double real radiation at tree level and 
single real radiation at one loop based on 
antenna functions. These antenna 
functions describe the colour-ordered radiation of unresolved 
partons between a 
pair of hard (radiator) partons. All antenna functions at NLO and NNLO 
can be derived systematically from physical matrix elements. 
To demonstrate the application of our new method on a non-trivial example, 
we implemented the NNLO corrections 
to the subleading colour contribution to $e^+e^- \to 3$~jets. 

An immediate application of the method presented here is 
the calculation of 
the full NNLO corrections to $e^+e^- \to 3$~jets~\cite{new3j}. 
The antenna subtraction method 
can be further generalised to NNLO corrections to jet production in 
 lepton-hadron or hadron-hadron collisions. In these kinematical situations,
the subtraction terms are constructed using the same antenna functions, but 
in different  phase space configurations: instead of the $1\to n$ decay 
kinematics considered here, $2\to n$ scattering kinematics are required, which
can also contain singular configurations due to single or double initial state 
radiation. These require new sets of integrated antenna functions, accounting 
for the different phase space configurations in these cases.

\section*{Acknowledgements}
This research was supported in part by the Swiss National Science Foundation
(SNF) under contracts PMPD2-106101  and 200020-109162 and
 by the UK Particle Physics and Astronomy  Research Council.

\end{document}